# A thin film source in a solid-state diffusion experiment: CoO on SrTiO$_3$

Qian Ma[a], Jan Erik Rybak[a], Natalie Jacqueline Ottinger[a], Timo Kassubek[b], Jörg Hoffmann[a], Karl-Michael Weitzel[b], Cynthia A. Volkert[a], Christian Jooss[a]

[a] Institut für Materialphysik, Universität Göttingen, Germany
[b] Fachbereich Chemie, Philipps-Universität Marburg, Germany

**Abstract**

To realize a chemical diffusion experiment for simple quantitative analysis of one-dimensional diffusion profiles requires the fabrication of a planar and chemically sharp interface between two phases, one serving as the diffusion source and the other as the material to be studied. We demonstrate a thin film source on top of single crystals or epitaxial films for the example of cobalt (II) oxide (CoO) grown on top of SrTiO$_3$ (STO) by ion beam sputtering. After deposition at room temperature, a nanocrystalline film with flat and chemically sharp interface is present. Diffusion annealing leads to a partial formation of the Co$_3$O$_4$ phase and recrystallization accompanied by a strong increase of the surface and the interface roughness. We report the conditions, where compact and stable CoO layers with flat interface can be maintained, serving as a constant source for Co diffusion. Exemplarily, the formation of a Co-diffusion profile is demonstrated after annealing of 240 h at 1163 K and comparatively studied by using three different methods: Energy dispersive x-ray spectroscopy (EDX) in a transmission electron microscope (TEM), atom probe tomography (APT) and time of flight secondary ion mass spectroscopy (TOF SIMS). Local and rather macroscopic concentration profiling do well agree within error.

**Introduction**

Because of their stability, chemical flexibility, and doping capability [1], oxide-based perovskites are of considerable interest for various application like catalysts [2] or thermoelectrical devices [3]. Whereas the mobility of oxygen can be rather high in some perovskite oxides used in solid oxide fuel cell applications [4], the mobility of cations in these materials is rather low, i.e., long-range diffusion mainly takes place at high temperatures [5]. However, the mobility of cations can be affected by tuning the vacancy concentration as well as by enhanced diffusion in dislocations or grain boundaries [6]. Charging of such extended defects can give rise to electric field-driven diffusion. In addition, the diffusion coefficient $D_c$ strongly depends on the point defect or cluster defect concentration that can give rise to a strong increase of $D_c$ [7].

The study of chemical diffusion requires the creation of a well-defined, chemically sharp interface between two phases, one serving as the source of the diffusing species and the other as the material system in which diffusion is to be studied. This then represents a well-defined starting point for diffusion studies via high temperature annealing by analyzing the spatial evolution of concentration profiles. For example, the determination of $D_c$ by using solution of the diffusion equation is easily possible in the case of a constant concentration source and a simple one-dimensional profile. This motivates the study of the phase formation and microstructural evolution of a thin film diffusion source that is grown on a single crystal surface during diffusion annealing.

We have chosen the diffusion of Co cations in SrTiO$_3$ (STO) as a model system for such studies. The diffusion couple was realized by depositing thin CoO films on almost atomically flat surfaces of single crystalline (100) SrTiO$_3$ substrates. Since the interfacial mixing is mainly determined by the deposition temperature $T_{dep}$ of the oxide film, we deposited the CoO films at room temperature by means of ion-beam sputtering. This method allows for preparation of dense and compact films also at such low deposition temperatures and dedicated layer configurations can be realized.

Investigations of one-dimensional diffusion profiles require stable planar interfaces and stable, approximately constant sources of the diffusing species. Since the thermally induced chemical diffusion of cations requires high post-annealing temperatures $T_{pa} > 1000$ K, various effects can change the concentration profile and lead to deviations from the one-dimensional profile. First of all, the existence of a fully covered chemically stable interface must be guaranteed, thus avoiding massive evaporation losses of Co.

However, cobalt oxides reveal two stable phases that can affect the vapor pressure: The monoxide CoO has a cubic crystal structure (space group $Fm\bar{3}m$) with lattice parameter 0.426 nm [8] and a melting temperature of $T_m \approx 1700$-$2100$ K that strongly depends on the oxygen concentration [9]. This phase is only stable at very low oxygen partial pressures. In air, CoO oxidizes to the spinel phase Co$_3$O$_4$ at temperature of about 870 K to 1070 K [10]. Depending on the oxygen partial pressure, Co$_3$O$_4$ (space group $Fd\bar{3}m$, lattice parameter 0.808 nm [11]) decomposes to CoO and O$_2$ at about $T \approx 1170$ K (in air) and $T \approx 1230$ K (at 1 bar O$_2$) [9], respectively.

In addition, microstructural changes affect the interface structure. Since room temperature deposition of a thin film typically results in a nano-sized grain structure, post-annealing will cause recrystallization. The evolving interface structure, in particular the crystal orientation of CoO might affect the transfer of the diffused species across the interface as well as possible changes of interface geometry due to the Kirkendall effect [12,13].

In this article, we analyze the annealing-induced phase and microstructural changes and demonstrate the successful realization of a CoO thin film as Co source for the study of Co diffusion in SrTiO$_3$ single crystals and SrTiO$_3$ epitaxial thin films. The concentration profiles are measured locally by application of Transmission Electron Microscopy (TEM) including analytical tools like Energy Dispersive Spectroscopy (EDX) combined with Atom Probe Tomography (APT) measurements. Furthermore, after removal of the recrystallized CoO layer by chemical etching, measurement of concentration profiles via Secondary Ion Mass Spectrometry (TOF SIMS) is achieved.

## Materials and experimental methods

We prepared CoO films with thicknesses of 120 nm and 760 nm on (001) oriented SrTiO$_3$ (STO) single-crystal substrates by ion-beam sputtering with a CoO target. During deposition, the partial pressures are $P_{Xe} = 1.0 \times 10^{-4}$ mbar (Xe as sputter gas), $P_{Ar} = 2.2 \times 10^{-4}$ mbar (Ar for charge neutralization) and O$_2$ with $P_{O2} = 1.6 \times 10^{-4}$ mbar. The deposition temperature $T_{dep}$ was fixed to room temperature.

In addition, homo-epitaxially grown STO layer ($T_{dep} = 1073$ K) was deposited on the STO substrate prior to the CoO deposition in order to study the effect of STO defect structure on the CoO properties as well as on the Co diffusion. The layers were deposited in situ by using a target changer. Note that the STO layer has Al as impurity that serves as a marker to differentiate the STO thin film from the single crystal.

The samples were post-annealed under different conditions. Here, we will focus on temperatures of $T_{pa}$ = 1123 K and $T_{pa}$ =1163 K in vacuum (background pressure 5×10$^{-6}$ mbar) and in Ar (background pressure 5×10$^{-6}$ mbar, $P_{Ar}$ ≈ 850 mbar at 1173 K). Note that post-annealing in Ar was performed in a closed system, i.e., the oxygen partial might be higher during the long-term post-annealing.

The as-prepared and post-annealed samples were characterized by means of x-ray diffraction (XRD, Bruker D8 DISCOVER), scanning electron microscopy (SEM: FEI NOVA NanoSEM 650) and cross-plane transmission electron microscopy (TEM, Titan 300 keV) equipped with analytical electron energy loss spectroscopy system (EELS: Gatan Quantum 965 ER image filter) and energy dispersive X-ray spectroscopy (EDX).

The electron transparent lamellas were prepared by means of focused ion-beam etching (FIB: Helios G4 Dual Beam system). This process requires Pt protection layers that are frequently visible in the TEM images. In order to minimize surface contaminations like redeposition of Co, the thinning direction was chosen parallel to the Co/STO interface and the lamellas were finally cleaned by Ar plasma etching (1 min.).

APT specimens were machined at the interface between CoO and STO using FIB machining via a conventional lift-out procedure [14]. The APT measurements were performed in a custom-built laser assisted wide angle atom probe with a 133 mm straight flight path and a conventional large diameter electrode, for details see [15]. The system was operated at a wavelength of $\lambda$ = 355 nm with a pulse length of 15 ps, a focus spot size of approximately $\omega_{1/2}$ = 150 μm, a repetition rate of 100 kHz and a base temperature of 120K. The detection rate window was set to 0.15 - 0.3 ions per 100 pulses. The data reconstruction and analysis were performed with the Scito 2.3.6 software.

Concentration depth profiles arising in the strontium titanate samples during annealing were characterized using time-of-flight secondary ion mass spectrometry (ToF-SIMS, IONTOF GmbH, Münster, Germany). The procedure used as well as the relevant SIMS parameters are identical to the conditions reported in [16] Bernzen (2024). In brief, the SIMS analysis has been performed employing a Bi$^+$ ion beam with kinetic energies of 25 keV (0.9 pA). Secondary ions are mass analyzed in a reflectron equipped with an extended dynamic range detector. Depth profiling is achieved by employing an O$_2^+$ sputter gun operated at 2 keV (80 nA). The area analyzed by the Bi$^+$ ion beam is 100 μm × 100 μm with a resolution of 128 × 128 pixels. Analysis and Sputtering are repeated until the desired depth in the material is reached.

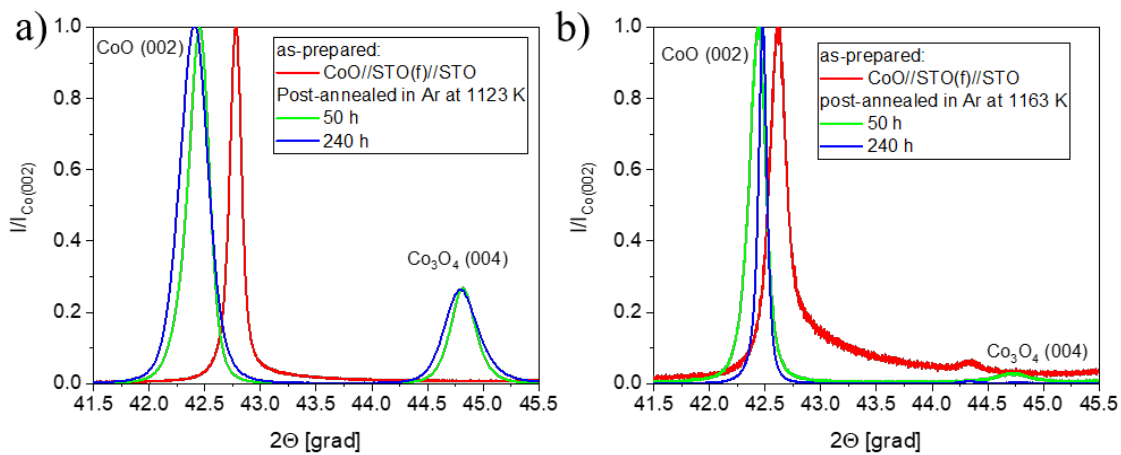

Figure 1: XRD Θ-2Θ pattern of as-prepared and post-annealed CoO film (shown in Fig. 1c). The intensities are normalized to the heights of the CoO (002) peaks. a) Post-annealed in Ar at $T_{pa}$ = 1123 K for 50 h and 240 h. b) Post-annealed in Ar at $T_{pa}$ = 1163 K for 50 h and 240 h.

# Realization of a CoO$_x$ thin film diffusion source on SrTiO$_3$

**Properties of as-prepared films:** The as-prepared cobalt oxide films are composed of the monoxide CoO with a preferred (001) growth direction. X-ray diffraction only reveals the (002) reflection of the CoO phase and no evidence for the Co$_3$O$_4$ (004) reflection (Fig. 1). That is in accordance with the results on cobalt oxide layers on Silicon and glass substrates, showing that IBS deposition at moderate temperatures and low oxygen partial pressures favor the formation of the CoO [17].

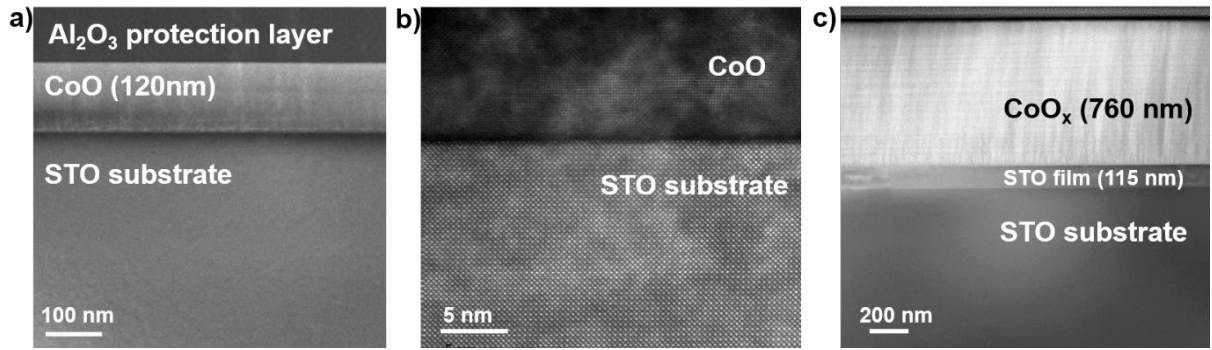

Figure 2: Cross-sectional STEM high-angle annular dark-field (HAADF) images of CoO as-prepared. a) Overview image of a CoO film on STO single-crystal substrate with an additional Al$_2$O$_3$ protection layer, deposited by IBS at room temperature. b) High-resolution image of a) near the interface to the STO substrate. c) Overview image of a CoO thick film, deposited on an STO film, homo-epitaxially deposited on an STO substrate at $T_{dep}$ = 1073 K

Figure 2 shows that the room temperature deposition of CoO results in dense and compact monoxide films on top of (001) oriented flat surfaces of SrTiO$_3$ single crystals (a,b) and homoepitaxially grown thin films (c). Independent of thickness and SrTiO$_3$ defect structure, the CoO layers are composed of small nano-sized grains with diameter of the order of 5 to 10 nm near the interface to STO (Fig. 2b) and columnar grain with diameters of a few 10 nm in the later growth states (Fig. 2a,c).

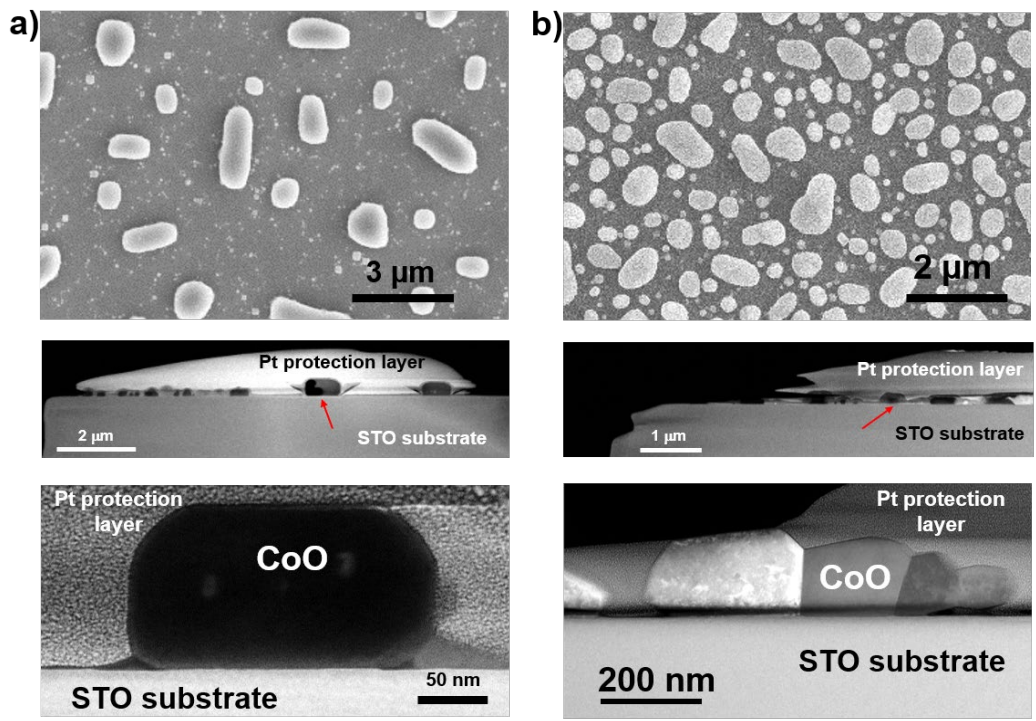

Figure 3: Planar view SEM image (top), cross-sectional STEM overview image (mid) and high-resolution image (botton) of a 120 nm thick CoO film after post-annealing at $T_{pa}$ = 1163 K for 20 h in vacuum (a) and Ar (b). The red arrows indicate the positions of the HR images.

The as-prepared CoO layers reveal smooth surfaces and interfaces, even for thick films of 760 nm, where the roughness is of the order of nm. The high-resolution image of the interface to STO (Fig. 2b) gives no evidence for preparation-caused crystallographic disorder or chemical intermixing. Spatial resolved line scans across the interface by means of Energy Dispersive Spectroscopy (EDX) and Electron Energy Loss Spectroscopy (EELS) shows that the Co signal decays within 2 nm (see also Fig. 5a) that is nearly the spatial and concentration resolution limits of the applied methods.

**Phase formation during post-annealing:** According to the equilibrium phase diagram [9], post-annealing at temperatures at low oxygen partial pressures should avoid the formation of the $Co_3O_4$ phase. But it has been pointed out that the stability ranges of the CoO phase are strongly reduced for nano-sized particles [18]. We indeed observe that post-annealing at $T_{pa}$ = 1123 K in Ar results in formation of significant amounts of the $Co_3O_4$ phase. The fraction of this phase does not strongly depend on the post-annealing time (Fig. 1a) but is substantially lower at post-annealing temperature of $T_{pa}$ = 1163 K (Fig. 1b). These results indicate that the oxygen partial pressure was high enough to enable the phase transition.

Oxidation of a compact CoO layer takes place via formation of a $Co_3O_4$ surface layer and precipitates within the CoO grains [19]. Reduction of $Co_3O_4$ nano-particles to CoO is only observed under high-vacuum conditions [20]. We therefore do not expect a reversible reduction reaction of $Co_3O_4$ to CoO during cooling. However, the post-annealing temperatures are close to the decomposition temperature of $Co_3O_4$ into CoO and $O_2$. This decomposition seems to strongly affect the volume fraction.

**Recrystallization of CoO layers:** In addition to phase changes, post-annealing leads to a pronounced recrystallization of the CoO layer. Depending on the initial film thickness, coverage of the STO substrate by recrystallized CoO layer is discontinuous or continuous.

Post-annealing of the 120 nm thick film at $T_{pa}$ = 1163 K in vacuum for 20 h (Fig. 3a) results in formation of CoO particles with heights of about 200 to 400 nm. These particles frequently reveal a well-defined shape and an orientation relation to the underlying STO substrate. Local diffraction pattern of such particles (Fig. 5b) show that the top facet (parallel to the substrate plane) reveal (001) orientation and the vertical facets are (110) oriented. The CoO(001) and (110) facets are connected by (111) facets (Fig. 3a, bottom). The different areas of the facets agree with the predicted anisotropy of the surface energies according to $\gamma_{100} < \gamma_{110} < \gamma_{111}$ [21]. The rather regular orientation of the formed particles (Fig. 3a, top) and the orientation of the facets implies that the [110] direction of CoO is parallel to the [100] direction of STO, i.e., an epitaxial recrystallization of the CoO layer. But we will exclude that some particles also reveal parallel [100] directions. This particle arrangement seems to be very stable. Increasing the post-annealing time to 50 h, only the thin CoO layer connected the large particles (see Fig. 3a, bottom) vanishes and the particle size increases (see also Fig. 5b).

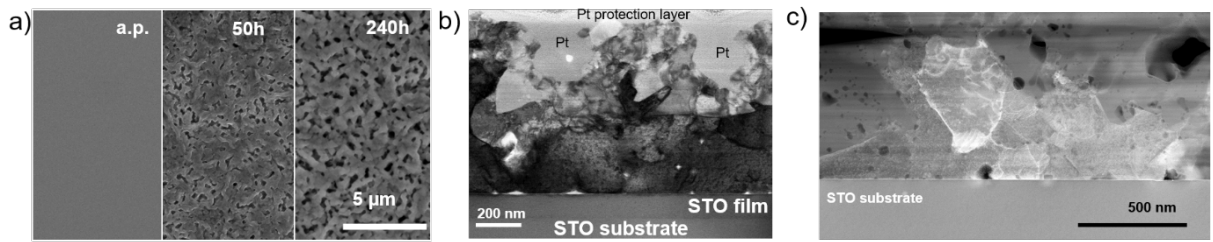

Figure 4: Post-annealing of a 760 nm thick CoO film shown in Fig. 1c at $T_{pa}$ = 1163 K in Ar. a) Planar view SEM image after different post-annealing times. b) Cross-sectional BF overview image after post-annealing for 240 h. c) STEM image of a CoO film post-annealed for 20 h.

The recrystallization is different, if the post-annealing is performed in Ar, but also results in a discontinuous covering (Fig. 3b). The height of the formed particles is smaller (between 50 and

200 nm) and the arrangement is less regular. The interface to STO reveal agglomerated particles without pronounced faceting with almost [100] CoO perpendicular to the substrate but no preferred orientation in plane.

In order to establish a constant concentration source, it is required to compare the volume of the deposited CoO film with the volume of the remaining CoO film after post-annealing. The latter can be estimated from the particle cross section and number density observed in SEM (top images in Fig. 3) and the particle heights deduced from the cross-sectional TEM investigations. Such comparison implies that post-annealing in vacuum cause a volume loss of about 60% for vacuum annealing and 30% for annealing in Ar.

Such evaporation losses might be affected by the size-dependence of the melting temperature because the grain sizes in as-prepared samples are of the order of 10 nm. But the losses might be also affected by formation of $Co_3O_4$ during post-annealing in Ar (Fig. 1), where the fraction of this phase is considerably lower at higher post-annealing temperatures.

**Evolution of surface and interface roughness:** A continuous coverage of STO by CoO is only observed for thick films. Figure 4a,b exemplarily shows the microstructural evolution of a 760 nm thick film forming an extraordinary large surface roughness after post-annealing for 240 h. However, even after this long-term annealing, the interface to STO is formed by a compact CoO layer. This layer consists of small grains with a [100] CoO out of plane texture (see appendix A).

The post-annealing conditions strongly affects the interface roughness. Strong oxidizing conditions result in a transition from flat pristine interface to a curved interface (Fig. 5a) which is not suitable for diffusion experiments. Reducing the post-annealing temperature and the oxygen partial pressure, the interface remains flat at post annealing times of 240 h (Fig. 5b,c).

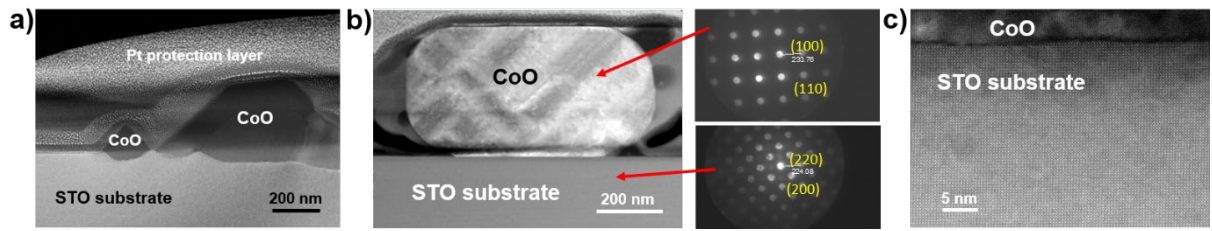

Figure 5: Cross-sectional STEM image of a 120 nm thick CoO film after post-annealing at 1270 K for 110 h in air. b) STEM image and diffraction pattern of a CoO particle formed after post-annealing at 1163 K for 50 h in vacuum. c) STEM HAADF image of the interface of the long-term post-annealed thick CoO film shown in Fig. 4c.

In summary, via the growth of CoO films of thicknesses of several 100's nm and post-annealing in Ar, a compact CoO layer is formed with a structurally stable and sharp interface, where chemical concentration profiles of cations can be measured using different analytical methods.

## Study chemical mixing profiles of cations and Co diffusion

In the following, we demonstrate the realization of a Co-diffusion profile in $SrTiO_3$ single crystals and epitaxial thin films by using different high-resolution analytical methods, i.e. EDX-TEM, APT and SIMS. All measured signals are normalized to their values far from the interface, respectively. The characteristic lengths of mixing are derived from the signal change from 90 % to 10 % of the maximum signal.

Figures 6a,b show the change of the Co signal across the interface between CoO and STO measured by EDX-TEM in cross-section lamellas. As-prepared (a), the Co signal decays within 2 nm which represents the spatial resolution limit. Long-term post-annealing of CoO on single-crystalline STO does not measurably change the concentration profile, i.e., the change is within the resolution limit. But by using the CoO layer as a source for diffusion into epitaxial STO

films, post-annealing strongly leads to broadened concentration profiles of Co, Sr and Ti (Fig. 6b). The decay length of all profiles in Fig. 6b is about 15 nm.

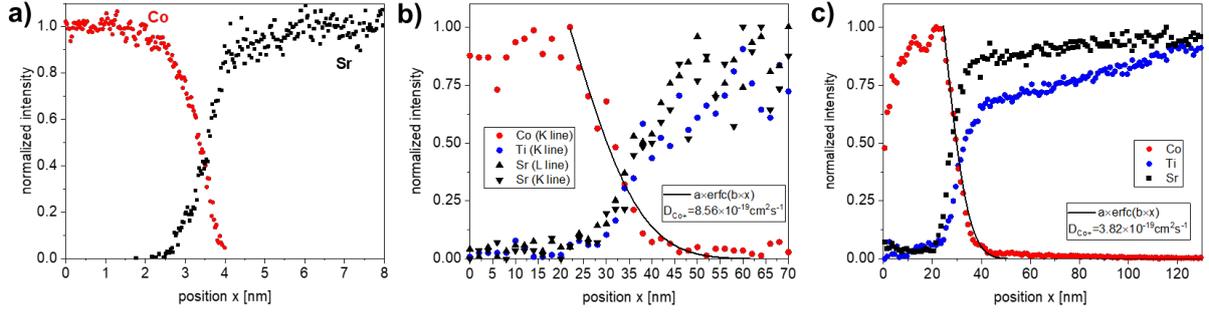

Figure 6: Determination of concentration profiles of cations across the CoO-STO interface by means of local methods. All concentrations are normalized to the intensities far away from the interface. a) EDX-TEM concentration profile across the interface of an as-prepared CoO on single-crystalline STO. b) EDX concentration profile of a CoO film on an STO film after post-annealing at $T_{pa}$ = 1163 K in Ar for 240 h. c) Concentration profile of the same film as in b) but measured by means of APT. For calibration and profile fitting, see text.

For an independent confirmation, we have also measured the diffusion profile in the same specimen by means of Atom Probe Tomography (APT) [15]. The result is shown in Fig. 6c. The position axis was calibrated by comparison with TEM images of the specimen. With this calibration, the Co decay length amounts to about 12 nm, comparable to the result deduced from EDX. We therefore conclude that the point defects in non-stoichiometric STO film strongly enhance the Co diffusion in STO.

The solution of the diffusion equation, the second Fick's law, for a semi-infinite body with a constant diffusion source of concentration $c_0$ is given by

(1) $\quad c(x,t) = c_0 \cdot erfc\left(\frac{x}{\sqrt{4D_c t}}\right),$

where erfc($x$) is denoting the conjugate gaussian error function and $x$ the coordinate perpendicular to the interface. A fit of the concentration profiles shown in Figure 6b and c gives for EDX measurement $D_{Co}$ = 8.6 ± 4.8 $10^{-19}$ cm$^2$/s and for the APT measurement $D_{Co}$ = 3.8 ± 0.9 $10^{-19}$ cm$^2$/s. Both measurements yield the same order of magnitude for $D_{Co}$ and are in the lower range of values observed for cation diffusion in SrTiO$_3$ thin films at 1163 K [7]. Here and in the following, the error bar for $D_{Co}$ is determined as an envelope function according to Eq. (1), which includes the scattering of the experimental data (see Appendix C). The obtained difference of $D_{Co}$ between the EDX and APT measurements is within this error.

**Removal of CoO layers:** The measurement limit and accuracy level of local concentration measurements by EDX-TEM is limited to about 1 at. % and by APT to between 0.001 and 0.1 at %, depending on the material. It is therefore desirable to combine such measurements with the high sensitivity method of secondary ion mass spectrometry (SIMS). This method is based on depth profiling, i.e., release of surface atoms and ions by ion-beam etching and subsequent analysis of the sputtered secondary ions. Due to the large surface area removed by the etching process, the sensitivity limit is in the $10^{-7}$ region. However, the depth profiles can by strongly modified by the surface roughness which would lead to significant broadening errors of the measured profiles.

In order to realize SIMS investigations of Co diffusion in STO, the post-annealed CoO layer with large surface roughness is removed by chemical etching in aqua regia. CoO is highly soluble in this acid, whereas only Sr-rich surface layers are removed in STO, thus aqua regia is used to generate Ti-O terminated STO surfaces [22].

The acid is composed of 25 ml distilled water, 50 ml HCL (32%) and 25 ml HNO₃ (65%). The basic step is a 5-minute etching in aqua regia followed by cleaning in distilled water in an ultrasonic bath. The sample is then cleaned with isopropanol. Depending on the thickness of the CoO layer, this procedure is repeated. The etching time depends on the actual surface roughness. In addition, compact layers of $Co_3O_4$ can hinder the chemical etching by means of aqua regia.

Figure 7b shows the concentration profiles measured by SIMS for an epitaxial STO film on STO with CoO post-annealing at $T_{pa}$ = 1163 K in Ar for 240 h (same as Figures 6b and c). For ex-situ ToF-SIMS analysis, the CoO layer on single crystalline STO was nearly completely removed by the chemical etching, as shown by the X-ray Photoelectron Spectroscopy (XPS) inspection in Fig. 7a. There are strong peaks of Ti 2p (shown in the inset). By taking Sr 3d into account (spectrum not shown), the quantitative analysis reveals a Sr:Ti ratio of 1 at the surface. In contrast, the weak Co 2p peak yields a surface-near Co concentration of 0.57 at%.

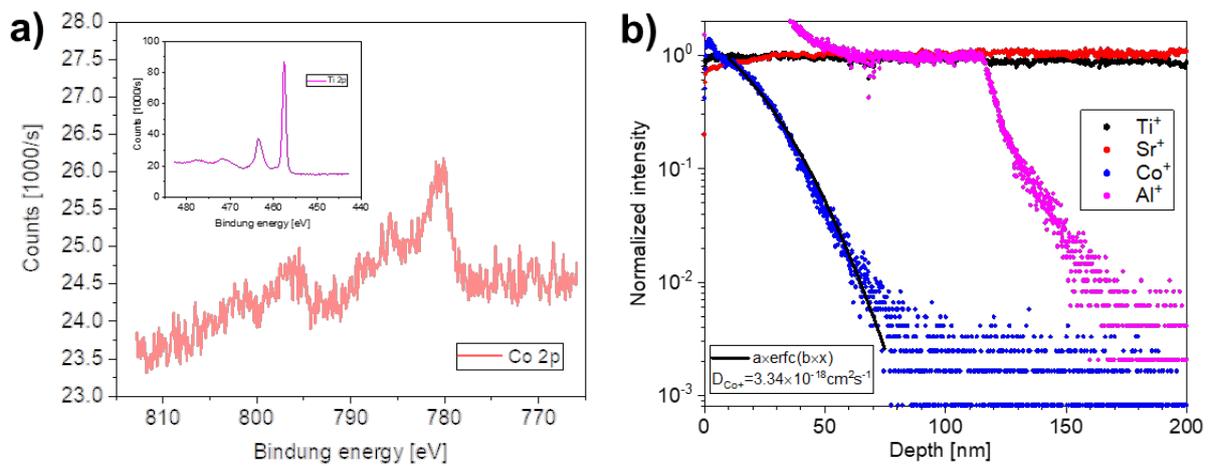

Figure 7: Chemical analysis after chemical etching. The samples were post-annealed at $T_{pa}$ = 1163 K in Ar for 240 h a) XPS spectra in the vicinity of Co 2p peak and Ti 2p (inset). b) Depth profiling by means of TOF SIMS. The intensities are normalized to their maximum values. The data includes the Al impurity of the STO film and the Co concentration profile was fitted by an error function (black line).

The concentration profiles shown in Figure 7b reveal a Co diffusion profile that starts immediately at the surface. Applying the same criterion for the decay length as used for the local measurements, the Co signal drops down from 90% to 10% within 34 nm. This is larger than the value deduced from local measurements (12 to 15 nm) but comparable in magnitude. The difference might be due to a small amount of CoO particles with a thickness of about 10-20 nm remaining on the surface after the chemical etching process, see appendix B. The related inhomogeneous sputtering during SIMS profiling can lead to an apparent broadening of the Co-profile.

Figure 7b shows depth profiles of 4 different elements, all normalized to 1 at a characteristic position. Sr and Ti are the main (cationic) constituents of the STO substrate and the STO film. The Co is the diffusing particle of interest. The Al is a trace element incorporated into the STO film during preparation of that film. Al is not contained in the STO substrate. The concentration profile of the Al impurity shows a plateau in the STO thin film and then sharply decreases in the single crystal. It thus can serve as a marker for the STO film thickness, indicating that Co-diffusion only took place in the STO film. Ultimately the thickness of the STO film derived above has been employed to calibrate the depth scale of the SIMS data. This is in good agreement with depth information obtained from tactile profiling of the ToF-SIMS crater.

The quantitative analysis of the Co diffusion profile by eq. (1) fit yields a Co diffusion constant in the STO film of $D_{Co} = 3.34 \pm 0.78 \cdot 10^{-18}$ cm$^2$/s at $T = 1163$ K. For fitting and error analysis see Appendix C. This value is around 5 times larger than the average value of $D_{Co}$ from EDX and APT. There are three possible sources of systematic error in comparing the concentration profiles measured by EDX/APT and by SIMS. One is the afore-mentioned presence of some nanoparticles of up to 20 nm height after etching, shown in appendix B, which will lead to a broadening of the SIMS profile. Since the surface coverage of such nanoparticles is only of about 15%, this effect is considered as minor. In addition, there might be a true spatial inhomogeneity of the diffusion profile due to point defect clusters in the ion-beam sputtered STO film. Furthermore, a possible impact of the lamella / tip preparation for TEM and APT measurements by means of Ga ion milling during FIB machining on the diffusion profile cannot be ruled out.

## Conclusion

Comparing different methods to create a diffusion couple, e.g. grinding and bonding, sandwiching thin foils between bulk systems, adsorption from the gas phase or film deposition (see the overview in [21]), it becomes clear that the major challenge is the control of the diffusion geometry and diffusion-induced microstructure. Analysis can become even more complex or even impossible, if diffusion-induced phase transformations are involved [22]. Approaches to growth films on a substrate are proposed e.g. in [21], however, to the best of our knowledge only rarely implemented (see e.g. [23]). The method of room temperature deposition of metal oxide films on top of single crystals or epitaxial films by means of ion-beam sputtering offers various advantages: The starting point before diffusion annealing is a dense, flat and compact film with surface roughness in the nm range and a chemical sharp interface and the emerging diffusion profile can be assumed to be parallel to the interface, resulting in a 1D diffusion profile. Challenges due to recrystallization, phase transformations in the layer-type diffusion source and evaporation losses give limiting conditions for post-annealing, however, can be to some degree controlled by film thickness and annealing conditions.

In the example of CoO on STO, the oxidation induced phase transformation and related increased evaporation losses can be reduced by performing the post-annealing of a rather thick film under Ar atmosphere instead of vacuum or air. The remaining $Co_3O_4$ formation can be avoided by post-annealing near the decomposition temperature of $Co_3O_4$. The selected post-annealing conditions also can have a huge impact on evolving interface structure via the Kirkendall effect, which needs to be avoided in order to realize simple geometries for diffusion profiling [24]. A strongly increased surface roughness due to recrystallization of the as-prepared nanocrystalline structure is almost unavoidable. Whereas this is not affecting the analysis of concentration profiles via local high spatial resolution methods, such as EDX-TEM or APT, a large surface roughness would strongly influence SIMS profiling. Thus, the rough film needs to be removed by suitable methods such as selective chemical etching, leaving the underlying single crystal intact.

Thin film deposition by ion beam sputtering additionally offers the possibility to tune the diffusion process in the studied material by growing thin films with a high concentration of point defects. The subsequent growth of the film that serves as diffusion source can be done in-situ by using different targets and a moveable mounting system. In the example of a homoepitaxial growth of STO thin films, the non-equilibrium preparation conditions can lead to Schottky point defect concentrations that are about three order of magnitude higher than the equilibrium concentration [25]. This offers the opportunity to study the impact of point defect concentration on the diffusion constant.

## Acknowledgment

This research project is funded by the German Science Foundation (DFG) as part of the research unit FOR 5065 ("Energy Landscapes and Structure in Ion Conducting Solids", ELSICS), projects P1, P4 and P6. The use of equipment of the "Collaborative Laboratory and User Facility for Electron Microscopy" (CLUE, Göttingen) is gratefully acknowledged.

## Appendix A

Figure A refers to Fig. 4c and shows selected area electron diffraction (SAED) patterns performed with a focused electron beam with smaller aperture (convergence angle 2 mRad) at different positions labeled by letters. The subgrains formed during recrystallization shows preferred orientations. Perpendicular to the substrate, they reveal an almost (001) direction and the in-plane (100) or (010) directions are almost parallel to the (110) direction of the STO substrate. The orientation relationship is analogous to that observed for well-separated particles after post-annealing of thin CoO films (Fig. 5b).

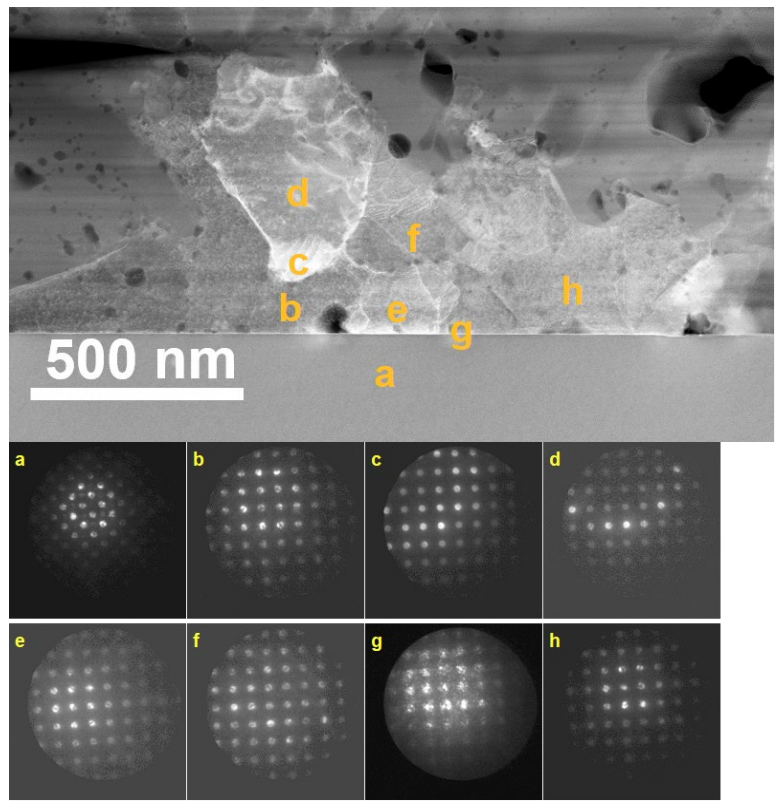

Figure A: Diffraction pattern of a thick CoO film post-annealed for 20 h in air. The letters indicate different positions a to h, where the diffraction was performed. The pattern a refers to the STO substrate.

## Appendix B

The chemical etching process leads to a drastic reduction in the surface roughness of post-annealed CoO films.

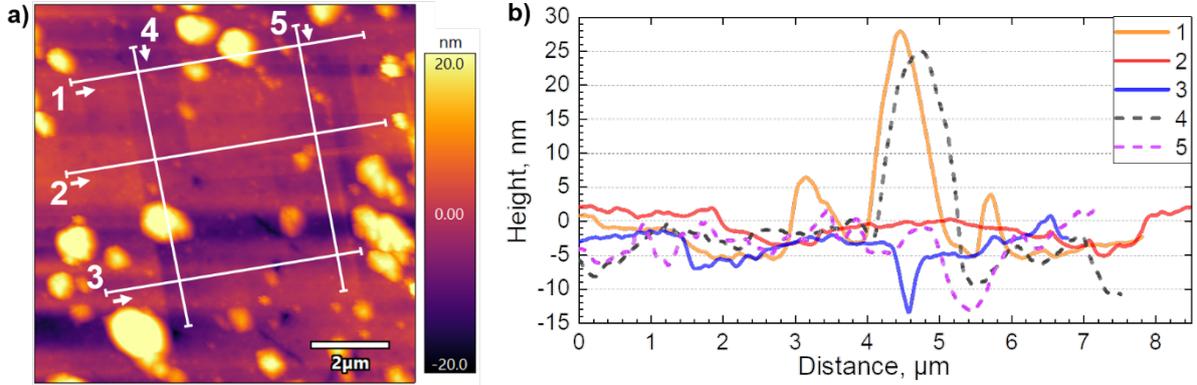

Figure B: a) AFM image of the surface of a chemical etched CoO film on single-crystalline STO. The sample was post-annealed for 240 hours at $T_{pa}$ = 1163 K in Ar. b) Line profiles along the lines 1 to 5 labeled in a).

Figure B shows an AFM image of the etched surface of a CoO film on a STO single crystal that was annealed for 240 hours at $T_{pa}$ = 1163 K in Ar. Examination by XPS (Fig. 7a) reveals an average Co concentration of 0.56 at% for this sample. The AFM image shows a remaining number of small particles with typical height of a few nm and a maximum height of 20 nm. These particles could consist of remaining (not removed) CoO and therefore contribute to a possible broadening of the concentration profile measured with TOF-SIMS (Fig. 7b).

## Appendix C

The conjugated error function

(C1) $\quad c(x) = a \, \text{erfc}\big(b \, (x - x_0)\big)$

gives reasonably good fits to the measurement Co concentration profiles. According to eq. (1), the diffusion coefficient $D_{Co}$ can be calculated by

(C2) $\quad D_{Co}(x, t) = \dfrac{1}{4(b \cdot x)^2 t}$,

where $t$ = 864 000 s (240 hours) is the post-annealing time.

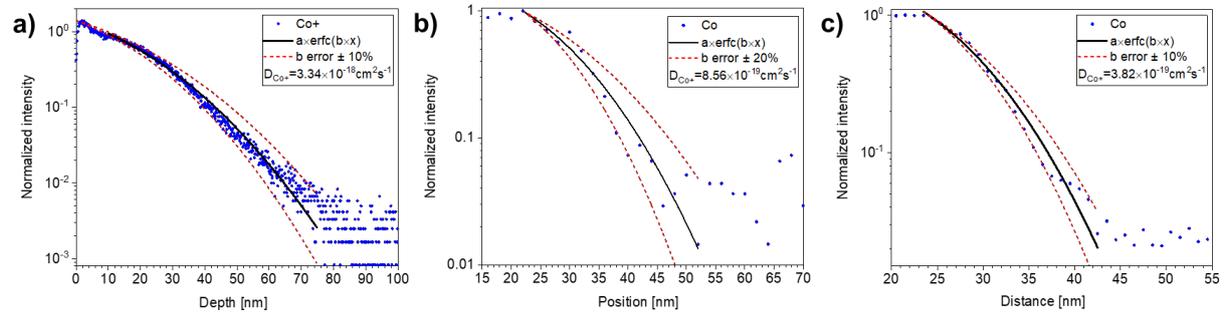

Figure C: Accuracy of the fit of Co concentration profiles for determination of $D_{Co}$. a) TOF SIMS depth profiling that refers to Figure 7b. b) EDX profile that refers to Figure 6b. c) APT profile that refers to Figure 6c. The solid lines show the obtained best fits and the broken lines represent the error bars for variations of the fit parameter b by 10% for TOF SIMS and APT, 20% for EDX.

Figure C refers to Figure 7b, Figure 6b and c, respectively, showing the accuracy of the fits. The range of variation of the parameter b is determined by the criteria that all data points of the

Co profiles are included in the range of maximum and minimum $b$. This leads to a variation of $b$ by ±10% for TOF SIMS and APT and by ±20% for EDX. The fit parameters, errors and resulting diffusion coefficients $D_{Co}$ obtained by the three methods are listed in Table C.

|  | a | δa | b | $D_{Co}$ (cm$^2$/s) | $δD_{Co}$ (cm$^2$/s) | δD/D (%) |
|---|---|---|---|---|---|---|
| **SIMS** | 1.39 | 0.007 | 0.029 | 3.34E-18 |  |  |
| δb = +10% |  |  | 0.032 | 2.76E-18 | 0.58E-18 | 17.4% |
| δb = -10% |  |  | 0.026 | 4.13E-18 | 0.78E-18 | 23.5% |
| **APT** | 1.06 | 0.013 | 0.087 | 3.82E-19 |  |  |
| δb = +10% |  |  | 0.096 | 3.15E-19 | 0.66E-19 | 17.4% |
| δb = -10% |  |  | 0.078 | 4.71E-19 | 0.90E-19 | 23.5% |
| **EDX** | 0.99 | 0.036 | 0.058 | 8.56E-19 |  |  |
| δb = +20% |  |  | 0.070 | 5.94E-19 | 2.62E-19 | 30.6% |
| δb = -20% |  |  | 0.047 | 13.38E-19 | 4.82E-19 | 56.3% |

Table C: Fit parameters a, b with errors and the corresponding diffusion coefficients $D_{Co}$.